# The structure of vortical flow over a rounded broad-crested weir

M. Anil Kizilaslan[a], Elif Deniz Atlas[b], Huseyin Yaban[b], Ender Demirel[b,*]

[a]Dept. of Construction, Canakkale Onsekiz Mart University, Canakkale 17020, Turkey.
[b]Dept. of Civil Engineering, Eskisehir Osmangazi University, 26480, Eskisehir, Turkey
*Corresponding author: edemirel@ogu.edu.tr
M. Anil Kizilaslan: makizilaslan@comu.edu.tr; Elif Deniz Atlas: atlaselifdeniz@gmail.com; Huseyin Yaban: hyaban02@gmail.com; Ender Demirel: edemirel@ogu.edu.tr

**ABSTRACT**

Turbulent flow over a rounded broad-crested weir is investigated by means of detached eddy simulation (DES) and large eddy simulation (LES) with special emphasis on the interaction of coherent vortex structures with free-surface. In order to set up and validate the computational model, experimental studies were conducted in a laboratory flume using a moderately rounded broad-crested weir with a rounding ratio of $R/P$=0.15, where $R$ is the radius of the upstream nose and $P$ is the height of the weir. The simulated mean velocity, Reynolds stresses and free-surface profiles show good agreement with the experimental measurements. Spatial variation of the boundary layer on the crest is well captured using a dimensionless form of the Lamb vector divergence. Boundary layer shape factor calculated over the weir was found to be lie between 0.76 and 0.92. A horseshoe vortex system emanating from the bottom of the channel interacts with the free-surface at the entrance of the crest causing undulation on the free-surface. Unsteady characteristics of the flow are examined in terms of the power spectral density (PSD) of vortex-induced forces acting on the weir. It is found that a free-surface boundary layer develops from the undulation to the wall boundary layer on the crest. It was revealed from the simulations for various Reynolds numbers that the installation of an artificial pool upstream of the weir significantly modified vortex structures and reduced undulation effects by 86% according to a proposed undulation index.

## 1. Introduction

Broad-crested weirs (BCWs) have been extensively used in past decades to measure flow discharge in natural streams and irrigation channels. Flow regime is significantly influenced by the existence of weir due to the local changes in turbulent flow. Separation of turbulent flow from the leading edge of the weir forms a recirculation bubble at the entrance of the crest, which may lead to diminution of flow discharge performance of the weir. Upstream corner of the weir is rounded in practice to reduce separation of the flow at the crest and sensitivity to silt deposition at the upstream face, as well as to increase flow discharge capacity of the weir. Interaction of large-scale vortical structures with the free-surface creates a complex flow structure over the crest, which is of critical importance for the development of a well correlated stage-discharge relationship. Therefore, investigation of the physical mechanisms that drive the flow over the weir is essential for academic research and practical applications.

Significant experimental studies were conducted in the literature to investigate correlation of discharge coefficient versus different dimensionless numbers consisting of weir dimensions and flow characteristics [1-4]. Azimi and Rajaratnam [1] suggested that surface tension effects could be neglected when the upstream head was larger than 30 mm. Boundary layer development on the crest causes the flow to decelerate due to the viscous effects. Hager and Schwalt [5] conducted experimental studies to examine the features of undular hydraulic jump on the crest of a sharp-edged BCW. Ohtsu et al. [6] found that flow conditions of undular hydraulic jump depended on the aspect ratio and inflow Froude number in a horizontal rectangular channel. Madadi et al. [7] reported that the upstream face of the BCW had a significant role on the undulation process and the undulation could be eliminated by changing the geometry of the upstream face. This particular feature of the BCW flow is the focus of the present study to account for the formation of undular flow at the entrance of the crest.

Sharp corner at the entrance of the weir caused the flow to narrow due to the recirculation zone created by the separation of turbulent flow from the corner. Entrance of the weir was rounded with a specific radius to reduce separation effects and hence to predict discharge accurately, which is essential for the application of water structures in irrigation channels. Flow separates at the upstream corner and a small recirculation zone forms at the entrance of the crest to reattach to the weir crest as a boundary layer. Vierhout [8] investigated the development of turbulent boundary layer on a rounded BCW and concluded that the boundary layer development on the crest was distinct from that on a flat plate in infinite fluid due to the strong interaction of boundary layer and free-surface. Boundary layer displacement thickness has been used in the literature as an alternative way to quantify characteristics of the boundary layer since definition of the boundary layer thickness on a curved surface is challenging. Moss [9] investigated the flow separation at the upstream edge of a square edged BCW using irrotational flow theory and concluded that the Laplace equation could give satisfactory results for the prediction of separation bubble. Boundary layer displacement thickness could be determined from measured flow velocities over the crest [8]. Radial pressure gradient and centrifugal force characterize boundary layer flows along the curved walls, and increasing external flow velocities produce an unstable shear layer over a concave surface with a rapid transition from laminar to turbulent boundary layer at the leading edge of the weir. Isaacs [10] further investigated the effect of laminar boundary layer on a BCW and reported that the boundary layer effects were significant on the estimation of energy losses, particularly at low discharges. Spatial variation of the boundary layer over the weir crest and the transition from laminar to turbulent flow near the rounded corner have been less understood due to the inherent limitations of experimental and analytical studies in the literature.

Developments in computer technologies in recent years have made it possible to analyze the internal flow structure of weir flows using Computational Fluid Dynamics (CFD) methods. Safarzadeh and Mohajeri [11] numerically investigated hydrodynamics of rectangular broad-crested porous weirs using RANS equations. Effects of the unsteadiness, non-hydrostatic pressure distribution and non-uniform velocity on the critical flow over the weir crest were investigated by Dai and Jin [12] using three models. Flow over a BCW with and without openings were investigated numerically for different slopes [13]. Armenio [14] emphasized that the LES has emerged as an effective tool for the investigation of the problems characterized by complex physics and geometry. CFD has emerged as a thriving method for the investigation of complex flow structures around water structures in the last decade. Interaction of turbulence with the free-surface were studied by [15-19]. Unlike the turbulence near the wall, tangential stresses vanish in the region adjacent to the free-surface and only the normal velocity component is restricted due to the kinematic condition. Restriction of the velocity component normal to the free-surface creates a thin layer adjacent to the free-surface and vanishing of the shear stresses forms a blockage-layer in the absence of wind [17]. Direct Numerical Simulation (DNS) based numerical studies conducted for the investigation of free-surface boundary layer used a flat free-slip plate assumption on the free-surface [17,18, 20] instead of considering deformations on the free-surface along the streamwise direction, which are significant for the present problem since the flow depth entering to the crest drops abruptly at the entrance, remains almost constant along the mid length of the crest and finally spills over the weir to the channel (Fig. 1). Although boundary layer development over the rigid wall of the crest has been considered in the development of stage-discharge relations, there is a lack of study on the effect of free-surface boundary layer.

Sketch of the turbulent flow over a round-nosed BCW is shown in Fig. 1 along with the dimensions and flow characteristics. Upstream corner of the weir is rounded with the radius of *R*. Flow enters to the flow domain at the inlet and emerges at the outlet of the channel with the same discharge when steady-state flow conditions are reached to maintain a constant upstream energy head. Here, *P* and *L* denote height and length of the weir, respectively, $h_0$ is the water depth at the inlet, *h* is the upstream water depth over the weir, and *Q* is the discharge. The main dimensionless numbers governing the flow over a BCW are Froude and Reynolds numbers, which are defined based on the flow conditions at the inlet as $Fr = q/\sqrt{gh^3}$ and $Re = h\sqrt{gh}/\vartheta$, respectively. Where, *h* inlet water depth excluding weir height, *q* is the unit discharge, *g* is the gravitational acceleration and $\vartheta$ is the kinematic viscosity. Weir is classified as BCW when the *h/L* is in the range of *0.08<h/L≤0.4* [5]. Round-nosed BCWs are classified according to the rounding ratio as slightly rounded for *0≤R/P≤0.094*, moderately rounded for *0.094<R/P≤0.25* and well-rounded for *0.25<R/P≤1.0*. In this study, experimental and numerical studies were carried out for a moderately rounded BCW, which has been extensively used in engineering applications since flow characteristics remain unchanged for well-rounded weirs [21].

Although numerical investigations of turbulent flow through a BCW have had great attention in recent years, large scale vortex structures induced by the channel bottom and weir have not been thoroughly investigated by means of well-resolved numerical simulations. Additionally, effect of the rounding on the flow characteristics has not been fully understood. Understanding of complex flow upstream of the weir and development of boundary layer over the crest with a rounded nose may provide insights to the design of the weir, which is the primary aim of the present study. A secondary aim is to reveal the boundary layer characteristics over the crest using high-resolution numerical simulations. Herein, the computational model was validated with the experimental measurements to simulate interaction of the boundary layer with the free-surface on the crest. The growth of the boundary layer, vortex structures, the interaction of the boundary layer with the free-surface and the mechanism that causes undulation effects on the free-surface are discussed in detail.

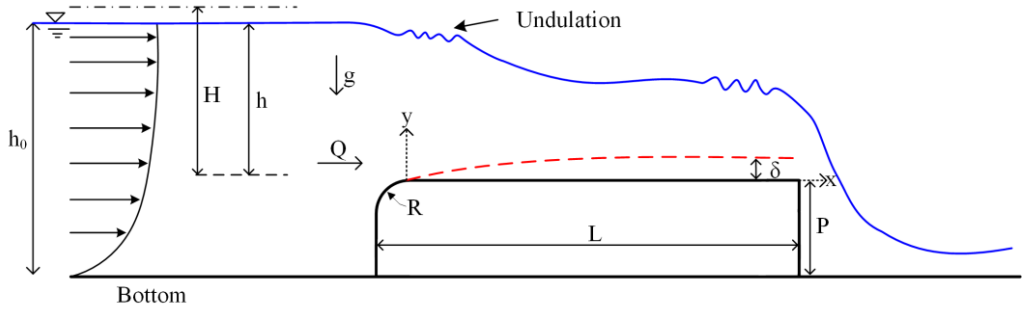

**Fig.** 1. Schematic view of the boundary layer flow over a round-nosed BCW with free-surface.

## 2. Flume Experiment

Experimental studies were conducted in a rectangular cross section laboratory flume having dimensions of 10 m length, 40 cm width and 60 cm height with horizontal bottom and glass walls. The flume bottom is stainless steel so that the model of the BCW can be fixed at a desired location using three magnetic hoists having the capacity of 250 kg. The discharge in the flume was adjusted manually using the valve at the inlet of the head tank and was measured precisely by the ultrasonic flow meter (Katronic KATflow 200) on the feeding pipe. The weir was located close to the downstream end of the flume to maintain fully developed flow conditions before the incoming flow reaches to the weir. Experimental studies consisted of measurement of velocity components and water surface profile at the upstream. Time-averaged velocities were measured using the Acoustic Doppler Velocimetry (ADV) at particular points (Fig. 2).

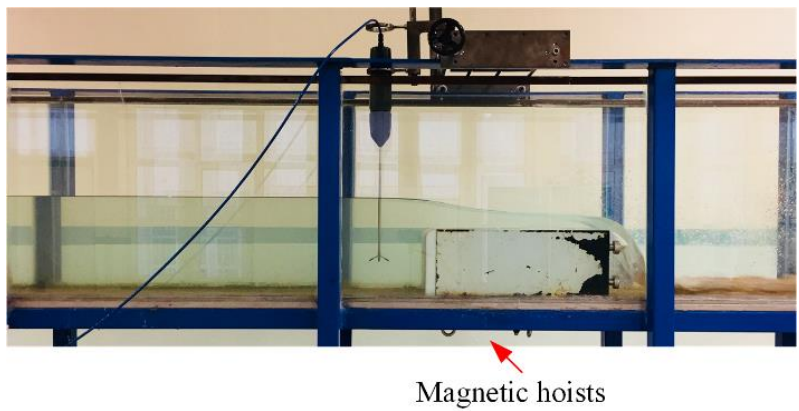

**Fig. 2.** Snapshot of the experimental setup.

Flow quantities at a point 5 cm away from the probes could be measured by using the ADV in three-dimensional manner. Preliminary experimental studies for different sampling durations selected from 60 s to 300 s revealed that statistically time-averaged values remain unchanged when the sampling duration is greater than 180 s. Measurements were conducted three times at a point and averaged of the measured data was used in order to reduce uncertainty arising from the experimental measurement setup. The sampling frequency rate of the ADV is 25 *Hz* and flow measurements were carried out during 180 *s* in order to collect adequate data to obtain time-averaged flow field [22]. Nominal Velocity Range (NVR), Sample Volume (SV) and Transmit Length (TL) parameters were set to 0.3 *m/s*, 0.6 *mm* and 1.3 *mm*, respectively to ensure that the signal to noise ratio (SNR) is greater than 20, which is recommended by the manufacturer. The transmit length and the sampling volume should be reduced to mitigate noise effects originating from the bottom while measuring flow velocities close to the wall. The highest measurement location is 5 *cm* below the free-surface. The raw data has been filtered using WinADV software [23] to exclude unphysical velocity measurements due to the noise effects in the flume.

Experimental studies were conducted on a weir model with dimensions of *P*=20 *cm*, *L*=53 *cm* and *R*=3 *cm*, which can be classified as a moderately rounded weir. As the discharge was adjusted to *Q*=20 *lt/s*, the water depth at the inlet was measured as $h_0$=29.5 *cm*. Thus, dimensions of the present weir are in the range of BCW flows (*h/L*=0.18). Photographs were taken using a video camera with 12 Megapixel and the frame rate of 30 *fps* when the flow reached to steady-state conditions. Position of the free-surface was captured by digitizing high resolution photograph taken during experimental studies.

## 3. Model Description

*3.1. Numerical Model*

A three-dimensional computational model is used for the in-depth investigation of turbulent flow over the weir including free-surface effects. The numerical model employs DES model, which is a hybrid of Reynolds-Averaged Navier-Stokes (RANS) turbulence model and Large Eddy Simulation (LES). The Shear-Stress Transport (SST) k-ω turbulence model is used as a RANS model in order to account for the boundary layer effects on the crest accurately since adverse pressure gradient, flow separation and possible reattachment of the flow on the crest may significantly affect flow structure. Lack of accuracy near the solid boundaries is the major drawback of an LES model and RANS models are not capable of calculating instantaneous vortex structures outside the wall region [24]. In order to mitigate such drawbacks, DES uses RANS model within the region close to the solid boundaries, whereas filtered LES equations are solved outside of this region depending on the turbulence length scale normal to the solid boundaries. Detailed description of the three-dimensional governing equations for the incompressible turbulent flow and k-ω SST DES can be found in Davidson [25]. DES aims to combine the most favorable aspects of the two techniques, RANS models for predicting the attached boundary layers and LES for the resolution of time-dependent and three-dimensional large eddies. LES is not capable of capturing relatively smaller structures that populate near the boundary layer [26, 27].

An accurate free-surface tracking method is required to capture undular flow over the crest. Free-surface position is calculated using Volume of Fluid (VOF) method [28], which is a robust free-surface tracking algorithm for the simulation of nonlinear surface waves. In the present study, numerical simulations are carried out using an open source CFD code OpenFOAM [29]. A specific boundary condition is used to maintain fully developed flow conditions at the inlet and incorporated into the solver to adjust constant head at the upstream of the weir, which will be discussed in the subsequent part of the study. First-order accurate Euler scheme is used for the approximation of time derivatives in the governing equations. Second-order upwind method is used for the advection fluxes and second order Gauss linear method is used for the viscous terms. Thus, the overall numerical scheme in the present study is first-order in time and second-order in space. A Flux-corrected Transport (FCT) based Multi-dimensional Limiter for Explicit Solution (MULES) method was used while solving the transport equation to guarantee boundedness and accuracy of the solution. The time step is automatically adjusted according to the Courant number to achieve a stable solution during unsteady simulations. Courant number is set to 0.5 in order to reduce truncation errors arising from the approximation of time derivatives. The time step size was found to vary around 0.00052 *s* during numerical solutions.

*3.2. Mesh and Boundary Conditions*

A block structured non-orthogonal mesh is adopted to capture weir geometry near the rounded corner precisely. As shown in Fig. 3a, blocks are constructed in the computational domain and numbered along the flow direction. Variable grid is clustered near the solid boundaries and free-surface to accurately calculate variations in flow variables close to the boundaries of water domain. A non-orthogonal mesh system is used in region 3 to capture upstream curvature precisely (Fig. 3b and Fig. 3c). Non-orthogonal correction is needed for the calculation of normal gradients near skewed faces. Thus, a second order non-orthogonal correction method with five steps is employed for the prediction of boundary layer effects precisely on the curved surface even if the non-orthogonal correction procedure has significantly increased simulation duration. Two-dimensional block-structured non-orthogonal mesh in the *x-y* plane is extended in the *z* direction to generate a three-dimensional mesh. Computational mesh is clustered near front and back faces of the computational domain for the accurate representation of contraction effects, which are significant for the flow over the BCW.

Four types of boundary conditions are defined in the present simulations, namely inlet, outlet, wall and free-surface boundary conditions. Synthetic fluctuations were produced for the turbulence kinetic energy and specific turbulent dissipation rate at the inlet depending on the mean velocity and turbulence intensity $(I = u'/U)$, which was calculated as 8% from the velocity measurements at the inlet. A laminar boundary layer is imposed at the inlet and boundary layer develops along the upstream channel from the inlet to the weir forming a turbulent boundary layer. Free-stream boundary conditions are applied for all flow variables at the outlet of the computational domain to prevent reflections of flow variables. Wall boundary conditions were applied for all variables at the surface of the weir. No-shear condition is applied for the velocity and zero gradient boundary condition is used for the pressure at the stationary walls.

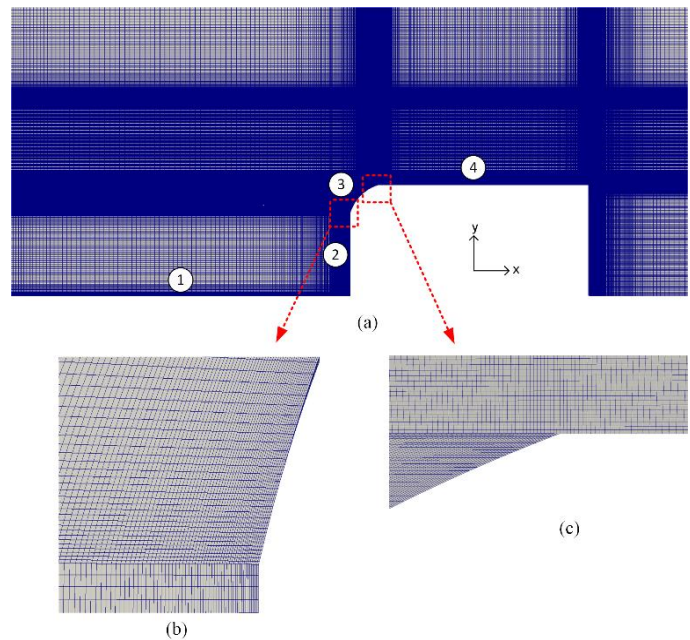

**Fig. 3.** View of the block structured mesh system: (a) General view of the variable mesh (unzoomed view) and block numbers, (b) zoomed views of the mesh at the upstream and (c) top of the rounded nose.

Unified wall functions are used for the turbulence kinetic energy, the specific turbulent dissipation rate and the turbulent viscosity as kqRWallFunction, omegaWallFunction and nutUSpaldingWallFunction in the OpenFOAM implementation ensuring that the first grid point neighboring to the wall is in the viscous sub-layer. A power-law velocity profile is imposed for the mean velocity at the inlet [30]:

$$U(y) = (\gamma + 1)U_0 \left(\frac{y}{h_0}\right)^\gamma \quad (1)$$

Where $\gamma$ is the shape factor, which is set to 0.1 for open channel flows, $U$ is the horizontal velocity component and $U_0$ is the average velocity at the inlet, which is calculated as the ratio of the discharge to the wetted area at the inlet section during simulation. This boundary condition is coded and incorporated into the original solver to couple inlet height and velocity field for a given discharge.

## 4. Results and Discussions

### 4.1. Time-Averaged Flow

Three-dimensional numerical simulations are performed for different flow rates given in Table 1. The numerical model is validated with the experimental data for *Q*=20 *lt/s*. The simulation was performed during 250 *s* to obtain statistically steady flow field in the flow domain. Time-averaging was initiated at t=50 *s* in order to exclude unrealistic fluctuations in the simulated flow originating from initial conditions.

**Table 1**

Flow rates in the experimental and numerical studies.

| Study | Design | *Q* (*lt/s*) |
|---|---|---|
| Flume experiment | NO URB | 20 |
| Numerical (1) | NO URB | 20 |
| Numerical (2) | NO URB | 10 |
| Numerical (3) | NO URB | 30 |
| Numerical (4 – 9) | URB | 20 |

Resolution of the computational mesh is critical for the accurate representation of boundary layer effects on the weir and undular flow on the free-surface at an acceptable time duration. Height of the first mesh adjacent to the wall has been carefully selected to calculate flow separation from the upstream nose and boundary layer development on the crest. Dimensionless cell sizes are calculated based on the average wall shear stress $(\tau_w)_{av}$ over the weir crest as $u_\tau = \sqrt{(\tau_w)_{av}/\rho}$. An extensive mesh independence study was carried out using five different mesh resolutions for the same flow conditions and weir geometry as in the experimental study to determine the required mesh resolution that yielded consistent results with the experimental data. Related dimensionless wall distances to the weir boundaries are calculated and listed in Table 2 for different mesh resolutions.

**Table 2**

Minimum, maximum and average values of dimensionless wall distances to the wall boundaries.

| Mesh | Cell Number | max $y^+$ value on the top of the weir | min $y^+$ value on the top of the weir | average $y^+$ value on the top of the weir |
|---|---|---|---|---|
| Mesh 1 | ≈ 1.5 million | 24.18 | 1.28 | 10.36 |
| Mesh 2 | ≈ 3.5 million | 18.28 | 1.02 | 7.9 |
| Mesh 3 | ≈ 8.5 million | 16.28 | 0.54 | 7.28 |
| Mesh 4 | ≈ 14.5 million | 14.9 | 0.39 | 7.06 |
| Mesh 5 | ≈ 22.5 million | 11.25 | 0.22 | 6.58 |
| Mesh 6 | ≈ 45 million | 10.7 | 0.18 | 6.32 |

Numerical simulations are performed for the mesh resolutions given in Table 2 and velocity profiles are compared in Fig. 4. Shear velocity was calculated using the average shear stress at the channel bottom to non-dimensionalize velocity and distance in Fig. 4. Simulation results are found to be independent of the mesh resolution for Mesh 4. Thus, this mesh is used for the rest of the simulations conducted in the present study. The computational mesh consisted of 14.23 million cells and maximum mesh non-orthogonality was calculated as 71.54°.

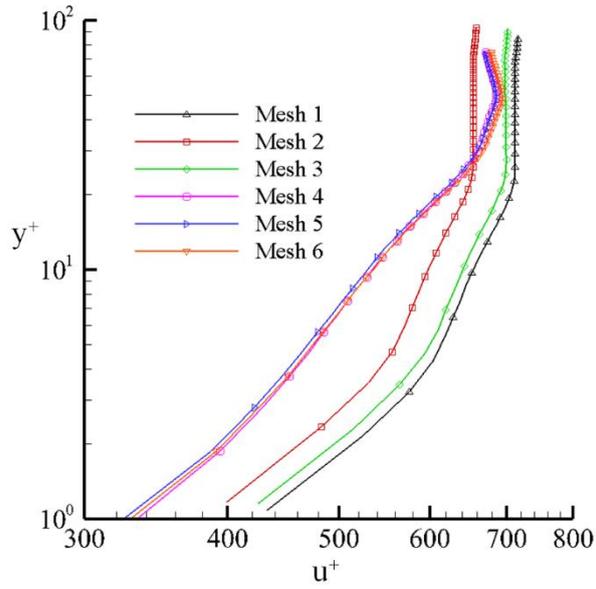

**Fig. 4.** Comparison of the velocity profiles at $x^+=10$ for different mesh resolutions ($Q=20$ *lt/s*).

Position of the free-surface was extracted from the numerical simulation results based on a threshold value of 0.5 for the volume fraction. Measured and predicted free-surface profiles are compared in Fig. 5. It can be observed from the figure that the present numerical model can accurately predict position of the free-surface at both entrance and outlet of the crest, since sudden changes were observed depending on the acceleration and deceleration of turbulent flow over the crest. Wiggles observed at the entrance of the crest were due to the undulation effects on the free-surface. As the wiggles propagate to the upstream face of the crest, a rapidly varied flow forms that can be characterized by the method of free-surface curvature [31]. This comparison also confirms that the present inlet boundary condition can accurately mimic inlet conditions for a specific discharge.

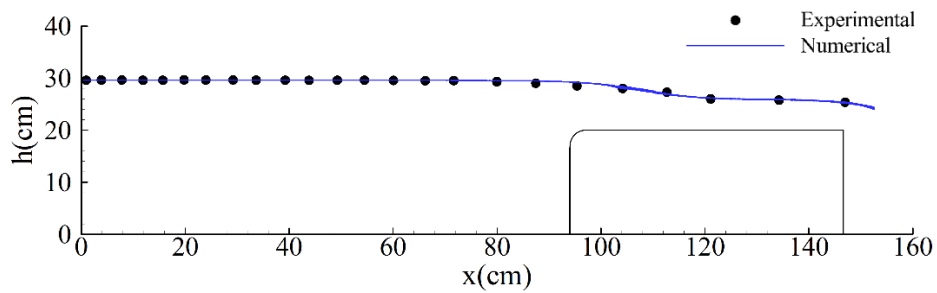

**Fig. 5.** Comparison of measured and simulated free-surface levels ($Q=20$ *lt/s*).

ADV measurements were carried out at 12 different stations upstream of the weir for $Q=20$

lt/s in order to obtain velocity profiles in the streamwise and spanwise directions. Coordinates were non-dimensionalized with respect to the inlet water depth ($h_0$) and the velocities were non-dimensionalized with respect to the averaged flow velocity at the inlet section ($U_0$), which was calculated from the discharge and inlet area. Measured and simulated mean streamwise velocities are compared at vertical and horizontal planes in Fig. 6a and Fig. 6b, respectively. Locations of the stations are shown in the figure as non-dimensional distances. Predicted streamwise velocities close to the free-surface increase as the flow approaches to the weir, which is found to be consistent with the experimental measurements. Flow velocities close to the boundary could not be measured due to the limitations of the ADV measurements. The numerical model can predict mean streamwise velocities close to the bottom, which is critical for the detection of the boundary layer development over the flume and weir crest. Predicted mean velocity profiles indicate that the fully developed flow at the far upstream of the weir transforms to the accelerating flow near the structure. As the flow approaches to the weir, backward velocities were observed close to the bottom due to the recirculation effects at the upstream of the weir [32]. The comparison of Reynolds stresses in Fig. 6c shows that the present numerical model can predict turbulence effects accurately. The discrepancy between experimental and numerical results observed close to the weir may be associated with the unsteady characteristics of the flow near this region.

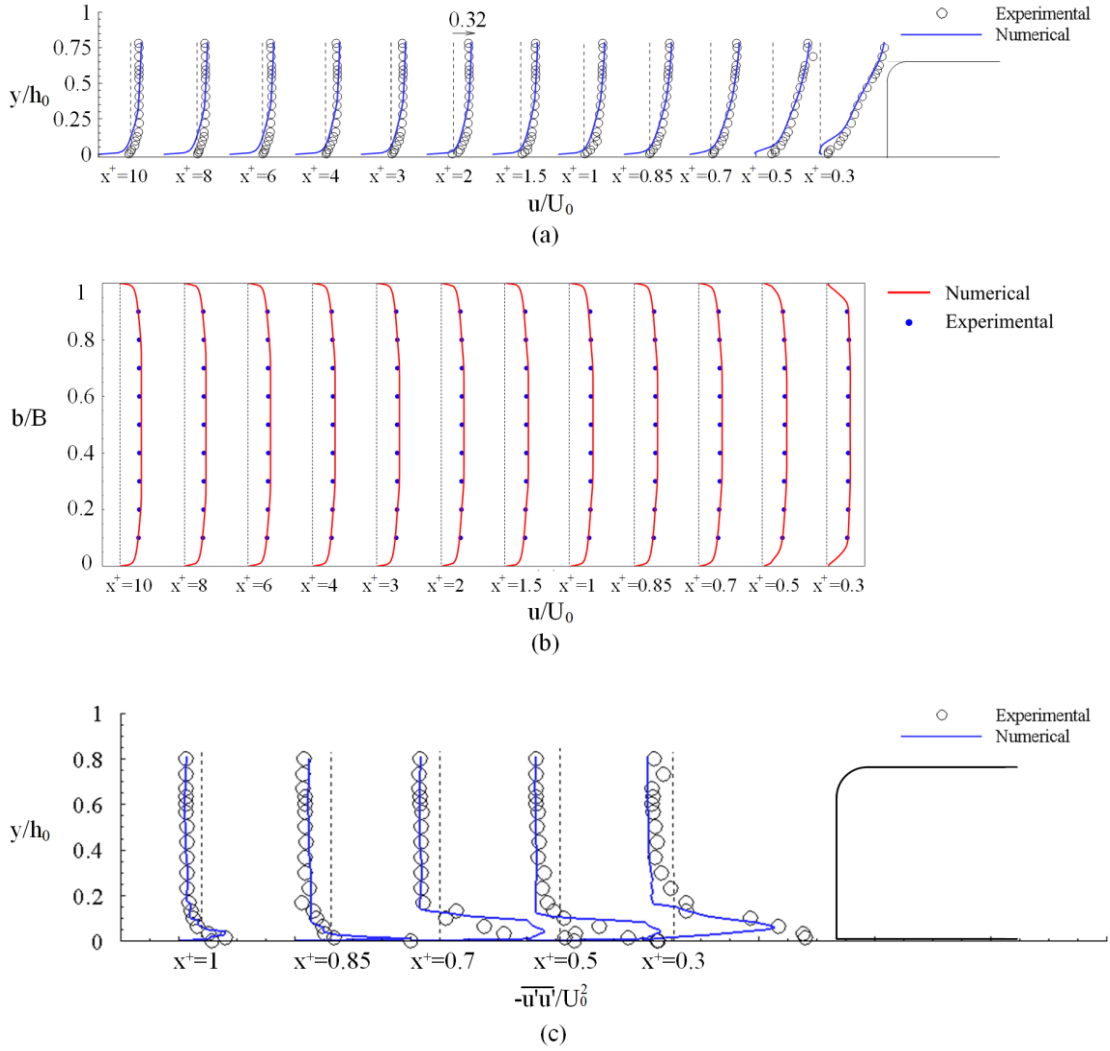

**Fig. 6.** Comparison of streamwise velocity components and Reynolds stress for experimental and numerical results ($Q=20$ *lt/s*). (a) Velocity profiles at the vertical plane ($z=b/2$), (b) at the horizontal plane ($y/h_0=0.5$) and (c) Reynolds stresses ($z=b/2$).

The relative error between numerical and experimental results are calculated using the following non-dimensional quantity at three measurement stations.

$$Relative\ Error = \frac{|U_{exp} - U_{num}|}{U_{num}} \qquad (2)$$

Here $U_{exp}$ and $U_{num}$ denote experimental and numerical velocities, respectively. Uncertainties are plotted in Fig. 7 for the relative error of 20% at different stations. It can be seen from Fig. 7 that acoustic waves scattering back from the walls to the flow region produce high uncertainty near the channel bottom and weir.

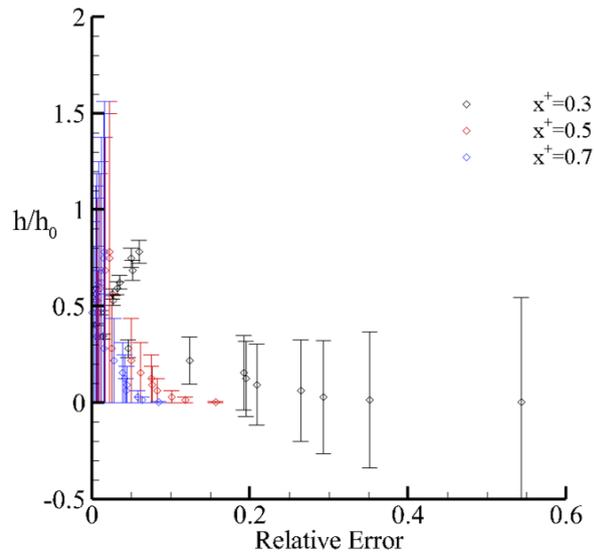

**Fig. 7.** Variation of the relative error along the vertical direction at different stations for 20% relative error.

Numerical simulation is performed under the same conditions as in the study of Imanian et al. [28] in order to further validate the present numerical model. Consistency between simulated free-surface profiles in Fig. 8a shows that the present numerical model can accurately capture deformation of the free-surface over the weir. Fig. 8b compares simulated velocity profiles with the experimental measurements of Imanian et al. [33]. The accelerating flow upstream of the weir is well captured by the present numerical model and results are consistent with the literature.

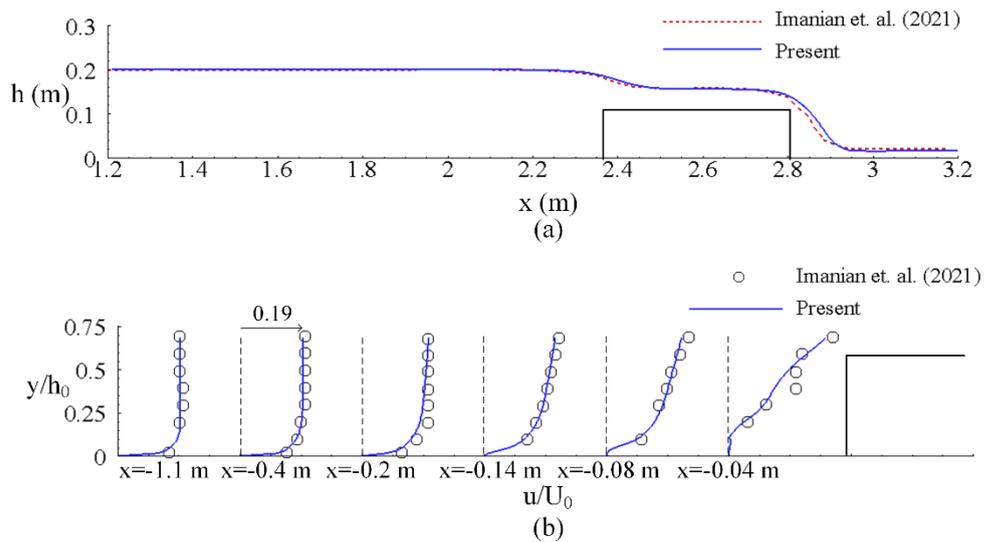

**Fig. 8.** Comparison of the numerical results with the literature data: (a) Free-surface profiles and (b) flow velocities.

*4.2. Coherent Vortex Structures*

Installation of a weir over the entire cross-section of the channel leads the turbulent flow to break down into different groups of vortical structures. Numerical simulation results are analyzed in terms of the spatial variations of time-averaged vortex-structures, which are interacted each other. In order to capture large scale coherent structures generated around the weir, *Q*-criterion [34, 35] is calculated from the mean velocity field. Fig. 9a presents three-dimensional vortex structures around the weir as the iso-surface of *Q*-criterion colored by the turbulence kinetic energy. Clockwise rotating (CWR) and counterclockwise rotating (CCWR) spanwise vortices develop starting from the separation point due to the Kelvin-Helmholtz instability and grow until they reach the upstream face of the weir, which forms a horn-like vortex (labeled CWR3 and CCWR3) elongated from the bottom corner to the leading edge of the weir. The formation of vortices advecting in a transverse direction plays an essential role in the onset of scour around the weir (Fig. 9b).

A horse-shoe vortex system (labeled HV1 and HV2) emanating from the bottom of the channel follows up to the upper layer of the recirculation zone and becomes elongated between free-surface and crest with a strong turbulence kinetic energy. An unstable shear layer evolves spatially in the streamwise direction at the entrance of the crest due to the occurrence of a strong horse-shoe vortex between free-surface and crest. Shear instability close to the free-surface produces streamwise vortices (labeled SW1) on the free-surface, which is observed as undulation in both numerical and experimental studies conducted in the present study. Streamwise vortices observed over the crest can be divided into two groups. The ones formed by the interaction of air and water in the area close to the free-surface are labeled as SW1 and SW2 (Fig. 9c), while the rib vortices formed above the weir are called as SWR1 and SWR2. Streamwise vortices observed at both entrance and exit of the weir are not only associated with local changes in the flow domain but also with the baroclinic torque caused by the high gradient of fluid density along the air-water interface [36]. Streamwise rib vortices (labeled SWR1 and SWR2) observed between the horse-shoe vortex and crest are responsible for the development of a mixed laminar and turbulent boundary layer on the crest [37], which will be further discussed in the subsequent part. LES captures vortex structures emanating from the channel bottom to the weir crest in Fig. 9(d-f). However, horn-like vortices observed in the DES results could not be captured by the LES, which may be due to non-linear boundary layer effects.

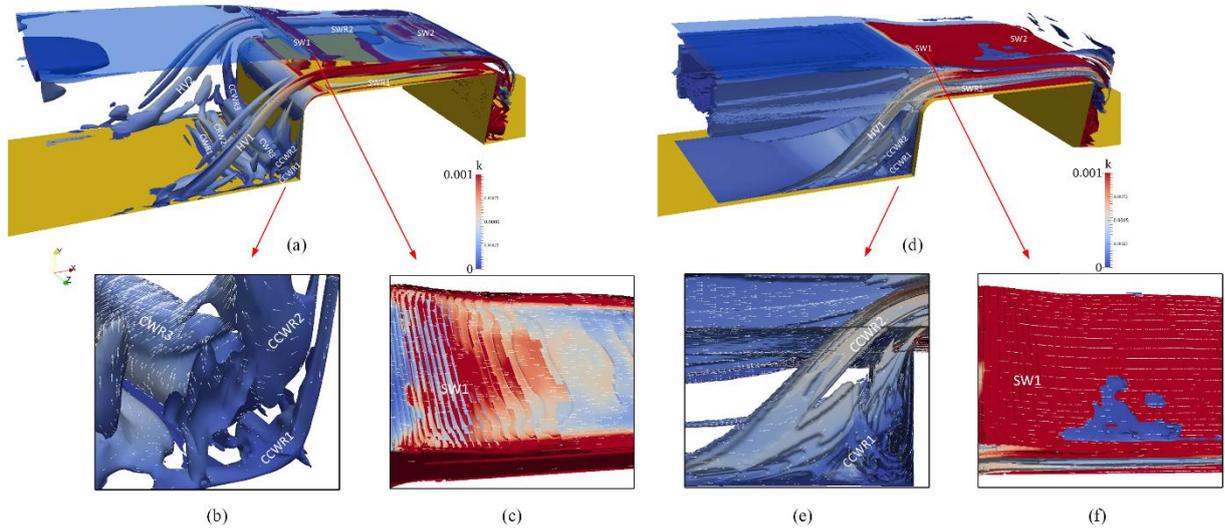

**Fig. 9.** Iso-surface contours of *Q*-criterion for *Q*=20 *lt/s*: Identification of different vortex systems forming around the weir for (a) DES and (d) LES, spanwise vortices upstream of the weir for (b) DES and (e) LES and vortex structures near the free-surface for (c) DES and (f) LES.

Numerical simulations are performed for *Q*=10 *lt/s* and *Q*=30 *lt/s* using the present numerical framework and results are compared in Fig. 10 to see the effect of discharge on the formations of vortex structures. A relatively weak vortices exhibit a scattered distribution of vortices around the weir when the discharge reduces and upstream water depth increases when the discharge increases and strong vortex structures form inside the recirculating flow upstream of the weir. The undulation observed on the free-surface becomes prominent when the discharge increases.

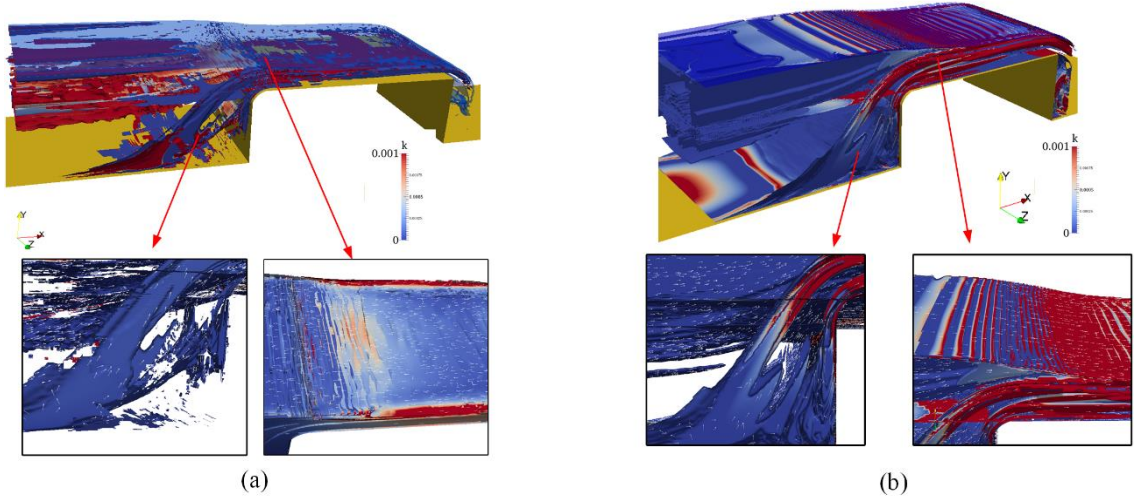

**Fig. 10.** Iso-surface contours of *Q*-criterion for (a) *Q*=10 *lt/s* and (b) *Q*=30 *lt/s* for DES.

*4.3. Boundary Layer Characteristics*

An understanding of boundary layer characteristics is important to the development of an accurate stage-discharge equation since boundary layer development over the crest is effective on the flow discharge capacity of the weir. The common method for the detection of the boundary layer is to find the normal coordinate to the wall at which the streamwise velocity component is approximately equal to the 99% of the free-stream velocity [31]. However, this approach is not applicable for the present problem due to fact that the flow velocity on the crest accelerates associated with the deformation of the free-surface in the streamwise direction, which makes it difficult to define a free-stream velocity over the crest. Furthermore, detection of the transition from laminar to turbulent boundary layer can be rather difficult or even impossible when free-stream velocity is used. In this study, the Lamb vector divergence [38] is used for the visualization of the interaction of different flow structures over the weir. The Lamb vector divergence is formulated as the scalar products of the divergence and Lamb vectors $\nabla \cdot L$, where the Lamb vector is defined as the cross product of the vorticity field by the velocity field as $L = \omega \times u$. The Lamb vector divergence is non-dimensionalized with respect to the local absolute value [22] to detect locations at which the sign of the Lamb vector divergence shifts:

$$\frac{\nabla . L}{|\nabla . L|} = \frac{u.\nabla \times \omega - \omega.\omega}{|\nabla . L|} \qquad (3)$$

Where, $u.\nabla \times \omega$ and $-\omega.\omega$ denote flexion product and negative enstrophy, respectively. Dimensionless value of the Lamb vector divergence is calculated from the simulated mean velocity and vorticity fields by incorporating above definition into the present open-source model. As seen in Fig. 11, the sign of the Lamb vector divergence switches between negative and positive at several locations due to the local changes in the flow. Laminar, transition from laminar to turbulent and turbulent regions in the boundary layer are well captured by the definition of the Lamb vector divergence in Fig. 11a. A laminar boundary layer developing from the intersection of the rounded nose and the crest turns into a transition boundary layer with the amplification of linear instability waves due to increasing local Reynolds number along the crest. Spatial variation of the boundary layer is extracted as the iso-surface with iso-value 0 of the Lamb vector divergence and the maximum height of the boundary layer is measured as 1.4 *cm* in Fig. 11b. The point where the boundary layer reaches 1.4 *cm* is given in Fig. 11b and the velocity profile at this location (*x*=30 *cm*, distance from the entrance of the crest) is shown in Fig. 11c. Undulation introduces a thin boundary layer elongated in the streamwise direction between free-surface and turbulent boundary layer. The surface layer is not observed for this problem since the thickness of the surface layer reduces with the Reynolds number [17]. While the flow motion vanishes in all direction on the rigid wall, only the velocity component normal to the free-surface is restricted by the kinematic condition on the free-surface.

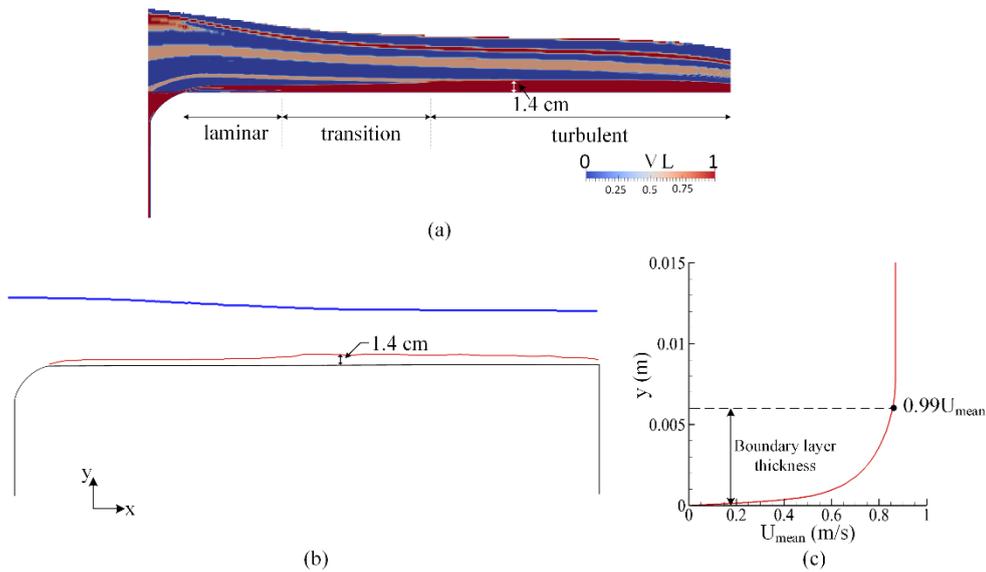

**Fig 11.** Representation of boundary layer characteristics over the crest for *Q*=20 *lt/s*: (a) Contour plot of the dimensionless Lamb vector divergence, (b) spatial development of boundary layer and (c) velocity profile at *x*=30 *cm*.

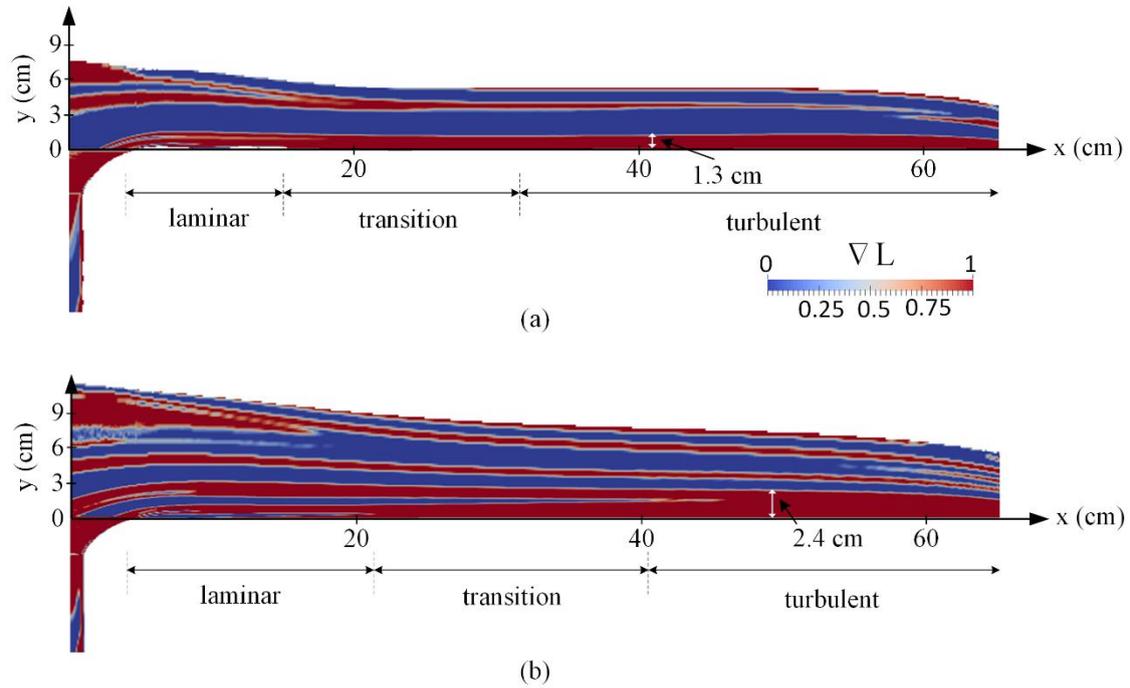

**Fig. 12.** Representation of boundary layer characteristics over the crest for (a) $Q=10$ *lt/s* and (b) $Q=30$ *lt/s*.

Fig. 12 shows the comparison of boundary layer characteristics for $Q=10$ *lt/s* and $Q=30$ *lt/s*. Thickness of the boundary layer and length of the laminar boundary layer increase when the discharge increases. Another important observation from this comparison is that local changes in the Lamb vector increase due to the strong vortices emanating from the bottom of the channel. Chen and Liu [39] also demonstrated that Lamb vector is a very convenient variable in determining the boundary layer.

Boundary layer shape factors are calculated by integrating velocity profiles from the crest to the boundary layer thickness (Fig. 12) at different stations over the weir and variations of the shape factors are compared in Fig. 13a for DES and LES. A high shape factor observed at the entrance of the crest is associated with the separation of the flow from the rounded nose. Then, the shape factor remains almost constant along the weir crest indicating that the turbulent boundary layer does not separate. The present DES and LES provided identical results in terms of boundary layer characteristics. However, LES produces smaller shape factors due to dissipation effects neat the solid wall. Reynolds stress shown in Fig. 13b indicate existence of the turbulent boundary layer over the weir crest.

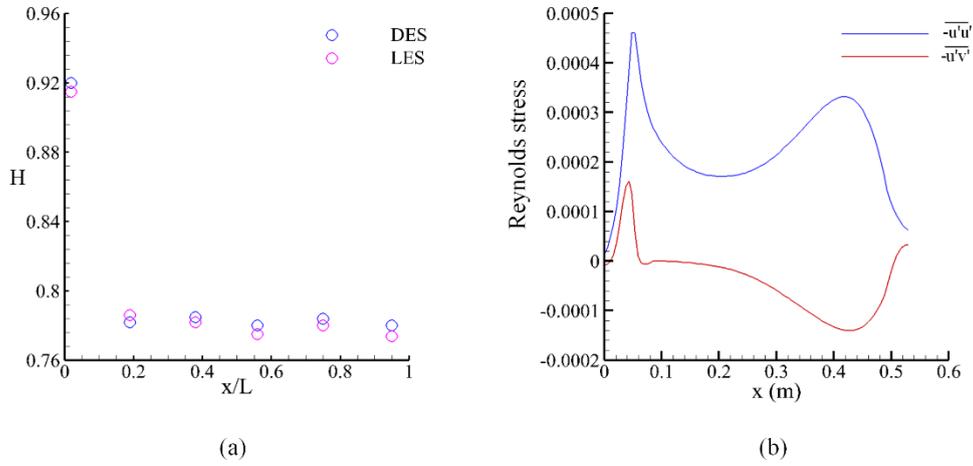

Fig. 13. Variations of the (a) boundary layer shape factors over the weir and (b) Reynolds stresses.

*4.4. Unsteady Characteristics*

Unsteady characteristics of the turbulent flow over the weir are investigated based on the experimental and numerical studies in this section. Velocity measurements were conducted at the probes located at the upstream of the weir (Fig. 14a) during 600 *s* and time variation of the velocities are shown in Fig. 14b. Probes 1 and 3 were located at a distance of *z/B*=0.1 from the lateral boundaries of the flume and probe 1 was located at the center of the section. Unsteady effects become prominent at the center of the channel due to the interaction of the recirculating flow with the incident flow. Present measurement setup can show the locations where unsteady effects are significant.

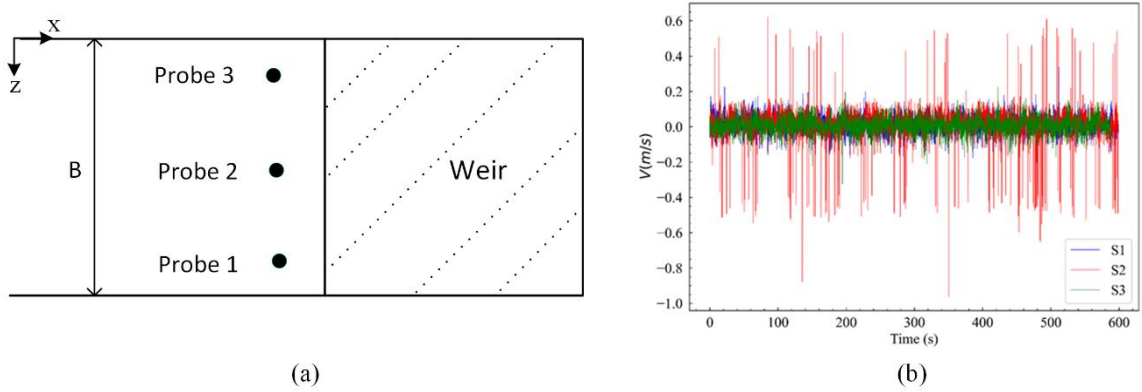

**Fig. 14.** (a) Measurement points at the upstream of the weir and (b) time variation of the velocity at the probes.

Three-dimensional streamlines are calculated from the mean velocity field and compared for RANS, DES and LES in Fig. 15. As seen in the figure, mean flow fields obtained from RANS and DES are not identical since energetic vortex structures captured by the DES alter mean flow field. The present DES implementation can capture the interaction of the recirculating flows with the incident flow better than the RANS model since RANS models can only consider time-averaged components of the turbulence fluctuations even for unsteady solutions. DES and LES capture unsteady vortex structures and produce an identical flow field upstream of the weir.

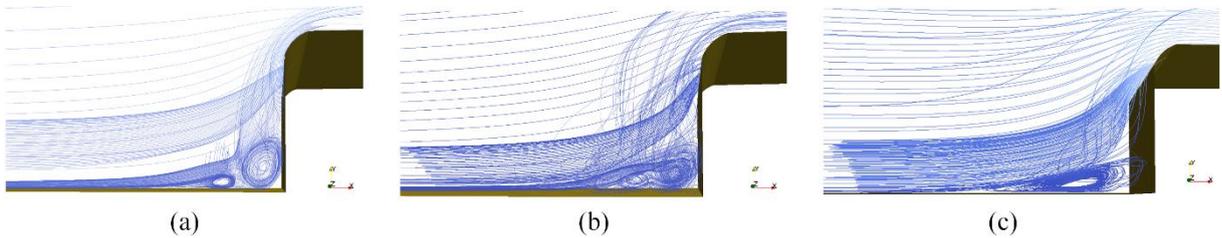

**Fig. 15.** Comparison of mean streamlines at the upstream of the weir for (a) RANS, (b) DES and (c) LES results.

Results were recorded with 1 s sampling frequency during numerical simulations and post-processed using the Paraview software, which is an open-source tool. We selected the time instants ($t=14$, $54$ $s$ and $69$ $s$) at which instantaneous vortex structures become prominent. Vortex structures are calculated from the instantaneous flow field of the DES at those time instants using the mean field as an initial condition and compared with the mean field in Fig. 16 for $Q=30$ *lt/s*. As can be clearly seen from the comparison that the flow over the weir is inherently dynamic due to the unsteady variations in the flow field, which can be predicted accurately using the present DES model.

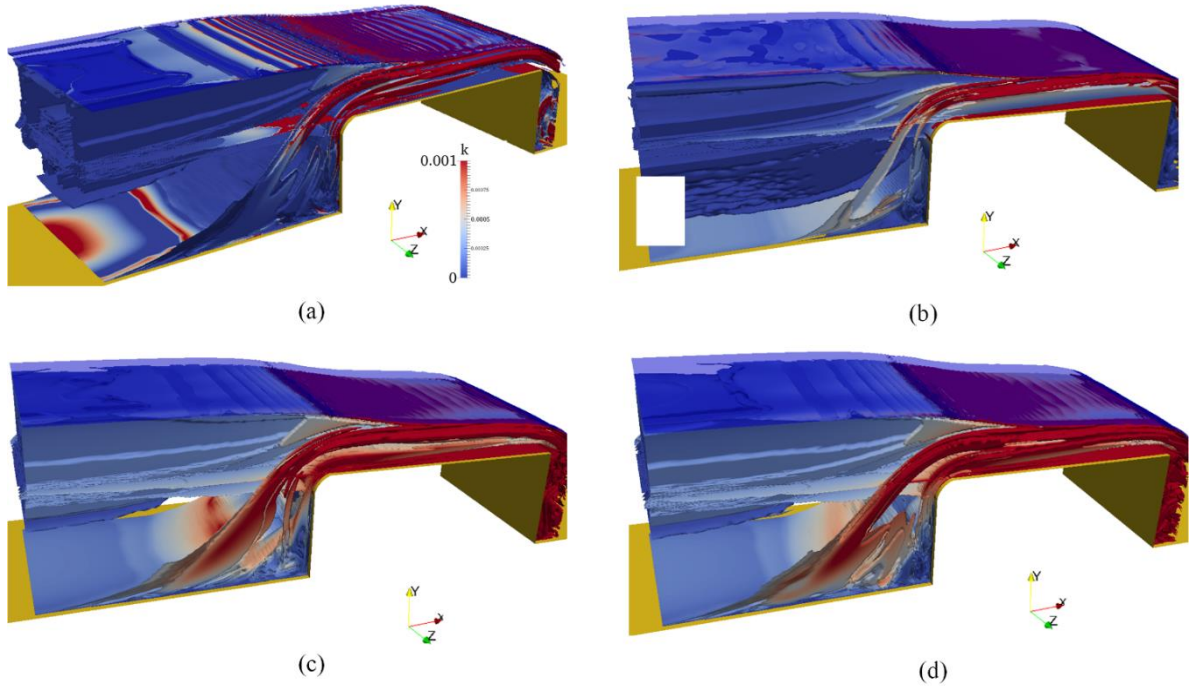

**Fig. 16.** Identification of different vortex systems forming around the weir with Iso-surface contours of *Q*-criterion for *Q*=30 *lt/s*: (a) mean, (b) *t*=14 *s*, (c) *t*=54 *s* and (d) *t*=69 *s*.

Viscous forces acting on the faces of the weir are calculated and recorded with a 100 *Hz* sampling frequency during the numerical simulation to see the effects of vortex-induced hydrodynamic forces acting on the weir. Power Spectrum Density (PSD) variations of viscous forces on the left face and crest are plotted in Fig. 17. Note that the values on the vertical axis are relative to a reference value. While viscous forces on the left wall of the weir are devoid of any significant energy level, viscous forces on the crest contain the peak energy at the frequency of $f \approx 5.9$ Hz. The present DES model can provide the spectral variations of viscous forces acting on the weir due the capability of capturing large scale vortices on the crest.

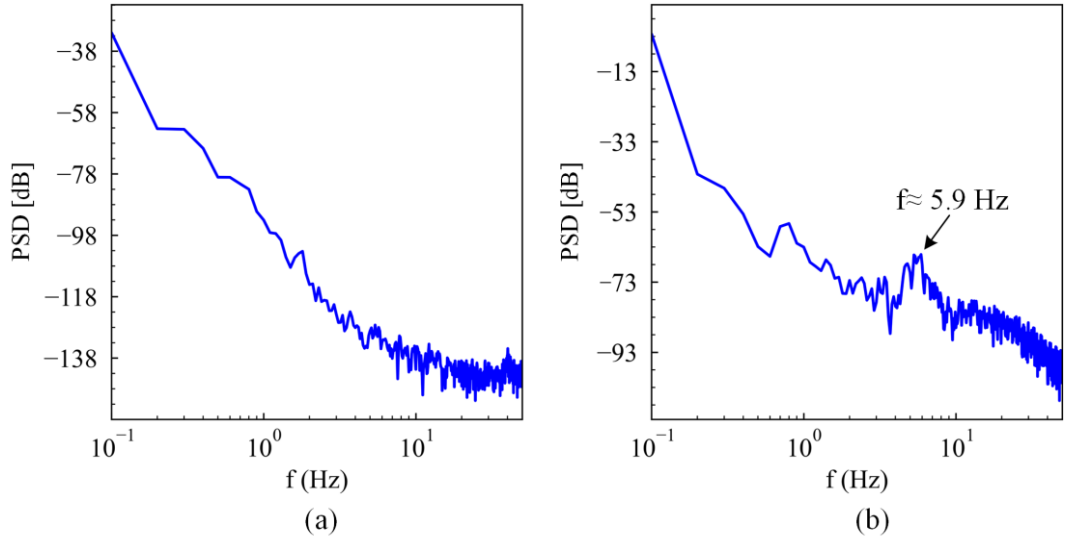

**Fig. 17.** PSD of the viscous forces on the (a) left face and (b) crest of the weir.

*4.5. Development of Undulation Reduction Basin*

As seen in Fig. 18, free-surface waves are observed as the flow approaches to the weir and undulation occurs at the entrance and outlet of the weir due to the formations of SW1 and SW2 (Fig. 9a), respectively. Formations of surface waves change near the midsection of the crest when the flow regime shifts from subcritical to supercritical spontaneously. Wave velocities at the center of the channel are observed to be greater than those near the solid walls in the subcritical region. However, wave velocity at the center of the channel increases in the supercritical region and an undulation is also formed at the outlet of the weir. This observation proves that undulation effects are prominent even for a weir with rounded nose and an alternative design should be used to mitigate undulation effects instead of changing the geometry of the upstream nose.

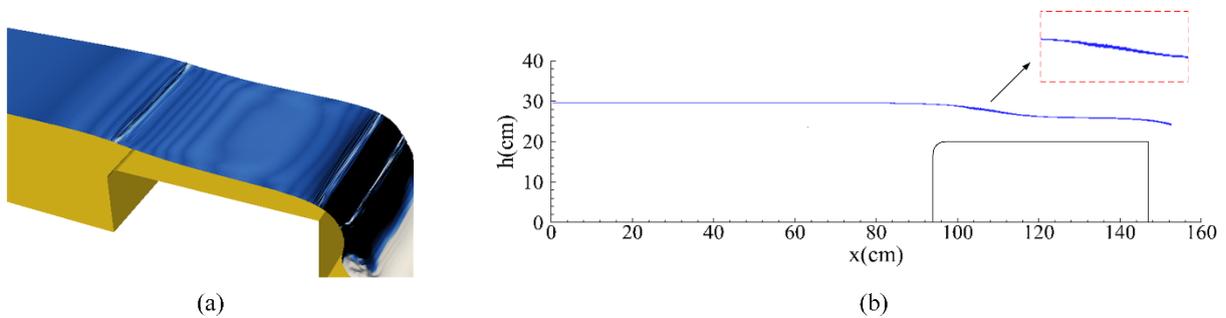

**Fig. 18.** Free-surface profile over the weir: (a) three-dimensional free-surface profile with undulation and (b) two-dimensional free-surface profile.

High-resolution numerical simulations conducted in this study reveal that the interaction of upstream vortex structures with the free-surface creates significant undulation effects at the entrance of a round-nosed weir. Based on the idea of confining upstream vortex structures to an artificial pool, an undulation reduction basin (URB) is designed based on the length of the upstream sloped face $L_s$, length of the bottom $L_b$ and depth of the pool $d$. As seen in Fig. 19, the URB is placed at a distance of $P$ from the upstream edge of the weir in an attempt to eliminate undulation effects. Free-surface is extracted from the simulations result and degree of the undulation can be calculated from the following undulation index (*UI*) for the assessment of the performance of the URB.

$$UI = \frac{1}{N}\sum_{i=1}^{N}\left|\frac{\partial^2 h}{\partial x^2}\right|_i \tag{4}$$

Where *N* is the number of points depending on the resolution of the computational mesh and *h* is the water depth at a point *i* on the free-surface. The use of second-order spatial derivative of water depth in this metric will exclude linear variations of water depth outside the undulating region. RANS simulations were performed using the same mesh and turbulence model, using different design parameters and calculated *UI* values are listed in Table 3 along with the undulation reduction ratios. It is seen that undulation reduction increases as the length and depth of the basin increase and the best performance is achieved for $L_b/P = 3$ and $d/P = 1.5$ among the design cases considered here. Despite the inclined pool leads the core of the recirculating flow to approach to the weir, performance of the URB with sloped face is lower than the rectangular design.

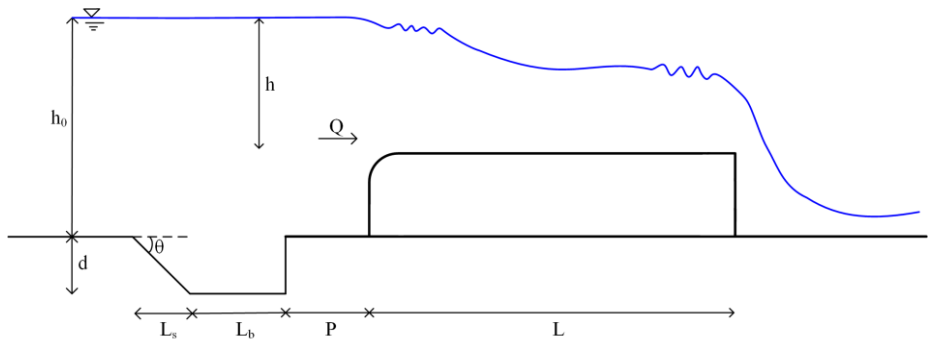

**Fig. 19.** Schematic view of the URB design.

**Table 3**

Design parameters of the URB for test cases.

| Design | $L_s/d$ | $L_b/P$ | $d/P$ | UI | Undulation reduction (%) |
| --- | --- | --- | --- | --- | --- |
| No URB | - | - | - | 0.29 | - |
| 1 | - | 2 | 1 | 0.108 | 62.76 |
| 2 | - | 2 | 0.5 | 0.106 | 63.45 |
| 3 | - | 2 | 1.5 | 0.0635 | 78.10 |
| 4 | - | 1 | 1.5 | 0.084 | 71.03 |
| 5 | - | 3 | 1.5 | 0.0606 | 79.10 |
| 6 | $\sqrt{3}$ | 3 | 1.5 | 0.1061 | 63.41 |

Visualization of streamlines at the upstream of the weir are compared in Fig. 20 for different designs. Strength of the recirculation zone observed near the upstream edge of the weir [40] reduces when the incident flow separated from the channel bottom is successfully trapped by the proposed URB design. Flow discharge performance of the weir is not influenced by the accumulation of the sediment upstream of the weir since the sediment can be trapped and settles within the proposed design, which is another contribution of the proposed design.

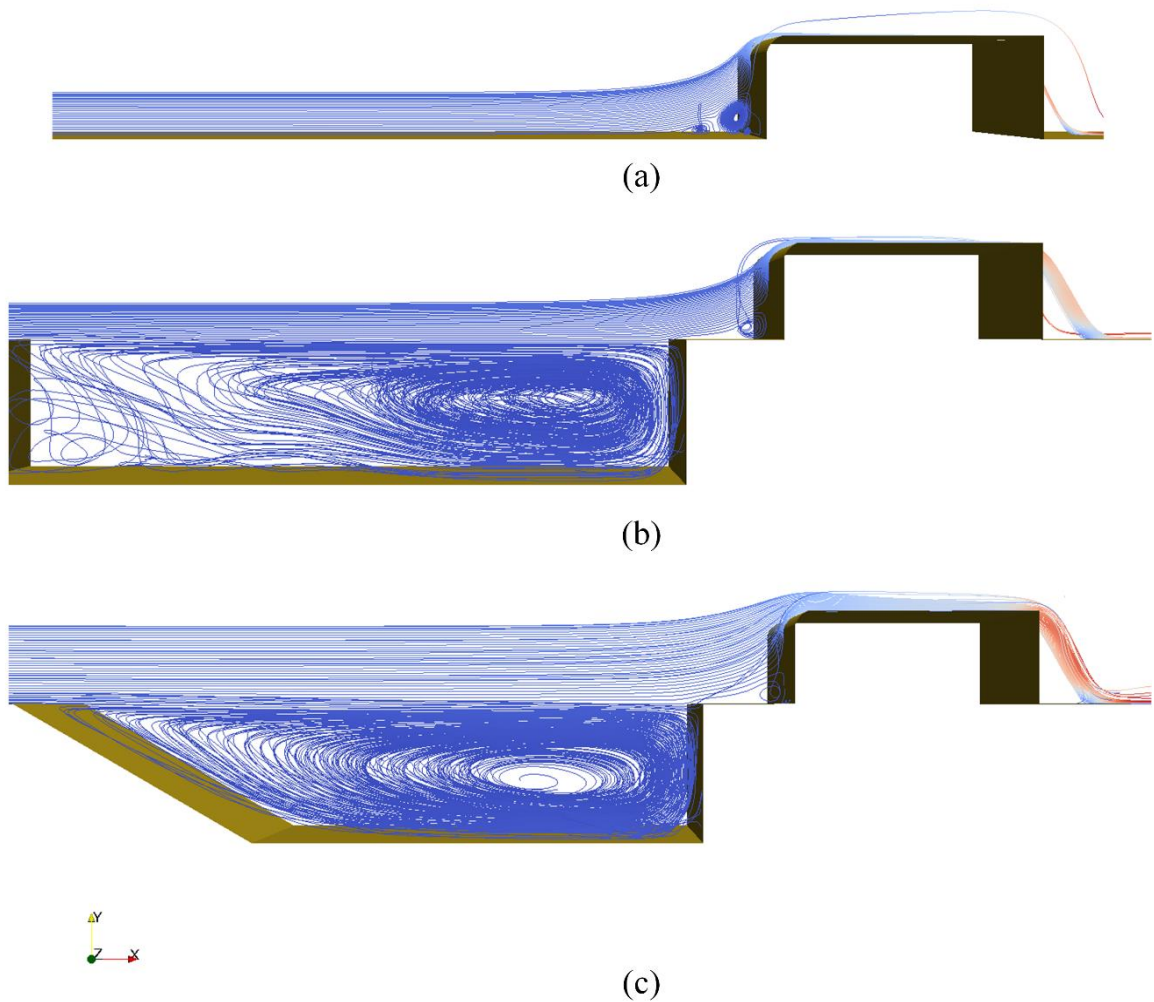

(a)

(b)

(c)

**Fig. 20.** Recirculation zones upstream of the weir with: (a) no URB, (b) URB (Design 5) and (c) URB (Design 6).

Simulations were performed for Re=1x10$^5$ and Re=1.6x10$^5$ on the URB 6 design to analyze the effect of the Reynolds number on the undulation reduction. The results given in Table 4 show that the reduction performance of the proposed design increases as the Re number increases. The proposed design can provide high performance in the reduction of the undulation for high Reynolds number flows in irrigation channel and streams.

**Table 4**

Design parameters of the URB for test cases.

| Design | Reynolds number | Q (lt/s) | UI | Undulation reduction (%) |
| --- | --- | --- | --- | --- |
| No URB | 130000 | 20 | 0.29 | - |
| 6 | 100000 | 10 | 0.128 | 55.87 |
| 6 | 130000 | 20 | 0.1061 | 63.41 |
| 6 | 160000 | 35 | 0.0405 | 86.04 |

## 5. Conclusions

A detailed numerical analysis of boundary layer flow over the crest of a moderately rounded broad-crested weir is provided using DES and LES in the present study. In order to confirm the accuracy and reliability of the numerical model, simulated flow velocities, Reynolds stresses and free-surface profiles were compared with the experimental data obtained in this study and in the literature. Numerical results are evaluated for identifying boundary layer development, vortex structures and wave propagation over the crest of the weir for different discharges. Present DES results revealed that spanwise vortices developed from the separation point on the channel bottom and lead to the formation of a horn-like vortex elongated from the bottom corner to the leading edge of the weir. Undulation was found to be created by the interaction of a horseshoe vortex system with the free-surface at the entrance of the crest. Dimensionless form of the Lamb vector divergence can be used to capture spatial variation of wall boundary layer over the crest and to determine boundary layer thickness and momentum thickness in free-surface flows. The boundary layer shape factor was found to lie between 0.76 and 0.92 over the crest, which indicates existence of the turbulent boundary layer over the weir crest. Separation of the flow from the upstream nose resulted in a sudden increase in the shape factor at the entrance of the crest. This result was also confirmed from the variations of the Reynolds stresses inside the boundary layer. PSD plots of the viscous forces acting on the weir crest show that round-nosed BCW experiences significant force fluctuations due to the dynamic structure of the flow over the weir. A free-surface boundary layer was found to be formed between free-surface and wall boundary layer resulting in a complex flow structure over the weir. Formations of free-surface waves propagating over the weir were observed to change at the critical section. A new design is proposed for the mitigation of undulation effects on the free-surface. A series of numerical simulations conducted for different design parameters and Reynolds number show that undulation can be reduced by 86% when the proposed design is used at the upstream of the weir. Another contribution of this design is that sediment can be trapped in the artificial pool during service life of the weir.

**Data Availability Statement**

The data that support the findings of this study are available from the corresponding author upon reasonable request.

**Acknowledgement**

This study was supported by the Scientific Research Project of Eskisehir Osmangazi University (Project No. 2017-1389). The numerical calculations reported in this paper were fully performed at TUBITAK ULAKBIM, High Performance and Grid Computing Center (TRUBA resources).